\documentclass[aps,pra,preprint,groupedaddress,showpacs,showkeys,floatfix]{revtex4}

\usepackage[dvips]{graphicx}

\newcommand{\bqn}{\begin{equation}}
\newcommand{\eqn}{\end{equation}}
\newcommand{\bqna}{\begin{eqnarray}}
\newcommand{\eqna}{\end{eqnarray}}
\newcommand{\bary}{\begin{array}{clcr}}
\newcommand{\eary}{\end{array}}

\begin{document}

\title
{Multiphoton ionization through the triplet states of Mg
by linearly and circularly polarized laser pulses}

\author{Gabriela Buica}
\affiliation{
 Institute for Space Sciences, P.O. Box MG-23, Ro 77125,
Bucharest-M\u{a}gurele, Romania}

\author{Takashi Nakajima}
\email[]{t-nakajima@iae.kyoto-u.ac.jp}
\affiliation
{Institute of Advanced Energy, Kyoto University,
Gokasho, Uji, Kyoto 611-0011, Japan}

\begin{abstract}
We theoretically study multiphoton ionization through the triplet states
of Mg by linearly polarized (LP) and circularly polarized (CP) fs
laser pulses.
After the construction of the atomic basis using the frozen-core
Hartree-Fock potential (FCHFP) as well as the model potential (MP) approaches
for both singlet and triplet series which show rather good agreements
with the existing data in terms of state energies and dipole matrix elements,
we solve time-dependent Schr\"{o}dinger equations with $3s3p$ $^{3}P_{1}$
as an initial state, and calculate the total ionization yield and
photoelectron energy spectra (PES).
\end{abstract}

% insert suggested PACS numbers in braces on next line

% insert suggested keywords - APS authors don't need to do this
%\keywords{Magnesium; Triplet states; Multiphoton ionization; Hartree-Fock;
% Model potential; Above-threshold ionization}

%\date{\today}
\maketitle

\section{Introduction}
\label{in}

During the last 30 years  many theoretical and experimental investigations have
been performed for Mg to obtain the atomic data and to understand its
interaction with radiation through {\it single-photon processes}.
The first extensive theoretical studies for the triplet states of Mg
were performed by Fischer \cite{fischer} with a multi-configuration
Hartree-Fock method which included correlations between the valence
electrons, and by Victor and   co-workers \cite{victor} with a semiempirical
model potential which included core-polarization and dielectronic terms
to calculate the oscillator strengths (OSs) for bound-bound transitions
with $^{1,3} S$, $^{1,3} P$, and $^{1,3} D$ symmetries.
Using a FCHFP with core-polarization and dielectronic terms
Chang \cite{chang1} calculated the OSs between
$3snl$  $^{1,3} L$ ($L=S,P,D,$ etc.) states of Mg.
Mendoza and Zeippen \cite{mendoza} studied photoionization from the excited
triplet state  $3s3p$ $^{3} P$ of Mg using a FCHFP with core-polarization
and dielectronic terms in the close coupling approximation.
Moccia and   co-workers \cite{moccia} developed a nonempirical
description of the core-polarization effects of Mg employing a basis set of
modified Slater-type orbitals to study the transitions between the
$3snl$  $^{1,3} L$ ($L=S,P,D$ , and $F$) states of Mg.
Luc-Koenig and   co-workers \cite{koenig} used an eigenchannel
\textit{R}-matrix and multichannel quantum defect theory (MQDT) to
investigate two-photon ionization of Mg atom.
Lately, Fang and Chang \cite{fang} studied single-photon ionization from the
excited singlet and triplet states of Mg below the Mg$^{2+}$ threshold
using an approach based on the \textit{B}-spline functions and Kim
\cite{kim} studied
single-photon ionization from the $3s3p$ $^{1,3} P$ states with a
\textit{R}-matrix
method combined with MQDT. Most recently Fang and Chang has developed a
\textit{B}-spline-based complex rotation method with spin-dependent
interaction to calculate atomic photoionization of Mg with singlet-triplet
mixing \cite{fang1}.

As for the {\it multiphoton processes} of Mg interacting with a laser
pulse there are several experimental and theoretical works, all of which
involve only {\it singlet states}:
Kim and   co-workers \cite{kim1} studied single and double ionization
of Mg by 10 ns Nd:YAG laser pulses at both 532 and 1060 nm
in the intensity range of $10^{12}- 10^{13}$ W/cm$^2$.
Druten and   co-workers \cite{druten} measured PES associated with
single and double ionization of Mg using 1 ps laser pulses in the wavelength
of 580-595 nm and $10^{12}- 10^{13}$ W/cm$^2$ intensity range, respectively.
Xenakis and   co-workers \cite{xenakis} investigated multiphoton
ionization of Mg using 150 fs laser pulses at the wavelength of 400 nm
for the peak laser intensities of up to $6 \times 10^{13}$ W/cm$^2$.
Gillen and   co-workers \cite{gillen2001,gillen2003}  measured the ionization
yield for single and double ionization of Mg exposed to the 800 nm, 120 fs
Ti:sapphire laser pulses for the peak intensities of
$10^{12}-10^{13}$ W/cm$^2$, which was followed by the theoretical analysis
\cite{gablamb}.
Liontos and his co-workers \cite{liontos} investigated single and double
ionization of Mg by Nd:YAG laser pulses with a ns duration for
peak intensities up to $10^{12}$ W/cm$^2$.
Zhang and Lambropoulos \cite{zhang} performed time-dependent calculations of
Mg for the case in which ions are left in excited states.
Recently we have studied the ionization yield and PES of Mg and clarified the
origin of the subpeaks in the PES by the second and third
harmonics of the fs Ti:sapphire laser pulse \cite{gabtaka2}.
Note that all the previous studies have focused on multiphoton ionization
from the \textit{singlet states} of Mg.

The purpose of this paper is to perform the theoretical study
for the multiphoton ionization processes through the {\it triplet
states} of Mg by LP and CP fs laser pulses.
Specifically we choose $3s3p$ $^{3}P_{1}$ as an initial state and perform
time-dependent calculations after the construction of the atomic basis
for both singlet and triplet series.
This paper is organized as follows.
In Secs. \ref{at} and \ref{se} we present the theoretical model:
The time-dependent Schr\"odinger equation (TDSE), which describes
the time-dependent interaction dynamics of the Mg atom with a laser pulse,
is solved on the atomic basis states of Mg with two-active-valence
electrons.
Atomic units (a.u.) are used throughout this paper unless
otherwise mentioned.
In Sec. \ref{nr} we present representative numerical results for the
state energies and the $J$-independent and dependent OSs between the triplet
states. Our results are compared with the existing data to confirm the
accuracy of our atomic basis.
Using those atomic basis states, we solve the TDSE
 to calculate the total ionization yield and PES from the
$3s3p$ $^3P_1$ initial state of Mg by LP and CP fs laser pulses.
Similar to the PES from the singlet ground state $3s^2$ $^{1}S_{1}$
of Mg \cite{gabtaka2}, the PES from the triplet  $3s3p$ $^{3}P_{1}$
state also exhibits subpeak structure.
Finally, concluding remarks are given in Sec. \ref{co}.

\section{Atomic basis states}
\label{at}

To start with, in order to study the interaction of the Mg atom with
a laser pulse  we have to construct the atomic basis of the Mg atom.
The Mg atom is a two-valence-electron atom; it consists of a
closed core (the nucleus and the ten inner-shell electrons $1s^2 2s^2 2p^6$)
and the two valence electrons.
As it is already mentioned in the literature \cite{bachau} there are
several approaches to solve the Schr\"odinger equation for
one- and two-valence-electron atoms in a laser field.
Since the general computational procedure has already been presented in
Refs. \cite{chang,tang,chang2} to construct the atomic basis states and
the specific details about the atomic structure calculation of Mg have been
reported in recent works \cite{gabtaka1,lambro}, we only briefly describe
the method we employ.
The field-free one-electron Hamiltonian of Mg$^+$, $ h_{a}(r)$, is
expressed as

\begin{equation}
 h_{a}(r)= -\frac{1}{2}\frac{ d^2}{{d}r^2}
           -\frac{Z}{r}+\frac{l(l+1)}{2 r^2} + V_{eff}(r),
 \label{eq:h_1e}
\end{equation}

\noindent
where ${r}$ represents the position vector of the valence electron, $Z$
the core charge, $l$ the orbital quantum number, and $ V_{eff}(r) $ the
effective potential acting on the valence electron of Mg$^+$.
Since the spin-orbit interaction is very weak for a light alkaline-earth-metal
atom such as Mg \cite{aymar,aymar1}, it might be safely neglected in the
atomic Hamiltonian for our specific purpose.
Similar to our recent study \cite{gabtaka1} in which we have presented
detailed comparisons between the frozen-core Hartree-Fock (FCHF) and MP
calculations for the singlet states of Mg, we employ two different
approaches in this paper to describe the effective potential,
$V_{eff}$, in Eq. (\ref{eq:h_1e}).
Namely (i) a FCHF potential and (ii) a MP.

\subsection{One-electron orbitals: Frozen-Core Hartree-Fock approach}

In the last years the most widely used method to describe the ionic core
is the FCHF approach. In the FCHF approach the effective potential
is given by

\begin{equation}
                V_{eff}(r)= V_{l}^{HF}(r) + V_l^p(r),
\label{v_effhf}
\end{equation}

\noindent
where $V_{l}^{HF}$ represents the FCHF potential  and
$V_l^p$ is the core-polarization potential which effectively
accounts for the interaction between the closed core and the valence
electrons \cite{chang}.
Specifically we employ the following form for the core-polarization
term:

\begin{equation}
V_l^p(r)= -{\displaystyle\frac{\alpha_s}{2r^4}
             \left[1-\mbox{exp}^{-(r/{r_l})^6}\right]},
\end{equation}

\noindent
in which $\alpha_s= 0.491$  is the static dipole polarizability of Mg$^{2+}$
\cite{mendoza} and $r_l $ ($l=0,1,2,...$) are the cutoff radii
 for the different orbital angular momenta:
$r_0 = 1.241$, $r_1 = 1.383$, $r_2 = 1.250$, $r_3 = 1.300$, and $r_4 =
1.100$ \cite{moccia1}.

\subsection{One-electron orbitals: Model potential approach}

Another simpler way to describe the ionic core is to use a MP,
$V_l^{MP}$ \cite{victor,preuss,aymar1,gabtaka1} instead of the FCHFP,
$V_{l}^{HF}$.
The advantage of the MP approach is that we can obtain the one-electron
orbitals without self-consistent iterations, since the interactions of the
valence electrons with the Mg$^{2+}$ core are replaced by pseudopotentials
for each angular momentum. Thus the complexity of the problem is greatly
reduced.
That is, instead of the FCHFP, i.e., $V_{l}^{HF}(r)$, we employ the
pseudopotential we have obtained in our previous work \cite{gabtaka1}
to describe the interaction of the valence electron with the Mg$^{2+}$ core:

\begin{equation}
V_l^{MP}( r) = V_{l}^{p}(r)
- \frac{A}{r} \exp{(-\alpha r^2)} + B_l \exp{(-\beta_l r^2)},
\label{MP}
\end{equation}

\noindent
where the values of the parameters introduced above, after the least-squares
fitting, are
$A  =  0.541, \; \alpha =  0.561, \;B_0 = 11.086, \;B_1 = 5.206,
\;  B_{l \geq 2} = 0, \; \beta_0 =  1.387, \;\beta_1 = 1.002,$
and
$\beta_{l \geq 2} = 0$ \cite{gabtaka1}.
We note that this form of $V_l^{MP}$ is different from the one
used in Refs. \cite{victor,preuss,aymar1}.
In Sec. \ref{nr}, we will compare the results obtained by FCHFP, MP,
and the experimental data.

In either approach described above to obtain the one-electron orbitals,
we employ a set of \textit{B}-spline functions to expand them.
Thus solving the Schr\"odinger equation for the nonrelativistic
one-electron Hamiltonian given in Eq. (\ref{eq:h_1e}) is now reduced to
an eigenvalue problem.

\subsection{Two-electron states }

Once the one-electron orbitals have been obtained using either the
FCHFP or MP, we can construct two-electron states with the configuration
interaction (CI) approach as we describe below:
The  field-free two-electron Hamiltonian,
$ H_{a}({\bf r}_{1},{\bf r}_{2})$, can be expressed as

\begin{equation}
\label{eq:h_2e}
H_{a} ({\bf r}_{1},{\bf r}_{2})= \sum _{i=1}^{2}
	       h_a(r_i) + V(\mathbf{r}_{1},\mathbf{r}_{2}),
\label{HF}
\end{equation}

\noindent
where $ h_a(r_i) $ represents the one-electron Hamiltonian for the
$i${th}  electron as shown in Eq. (\ref{eq:h_1e}), and
$ V(\mathbf{r}_{1},\mathbf{r}_{2}) $
is a two-electron interaction operator, which includes the static
Coulomb interaction $ 1/|{\bf r}_1 - {\bf r}_2| $ and the effective
dielectronic interaction potential \cite{chang,moccia}.
${\bf r}_1$ and ${\bf r}_2 $ are the position vectors of the two valence
electrons.
By solving the two-electron Schr\"odinger equation for the Hamiltonian
given in Eq. (\ref{HF}), the two-electron states are constructed
with the CI approach \cite{chang,tang,chang2}.
For Mg, which is a light alkaline-earth-metal atom, the $LS$ coupling is
known to give a good description and hence it is sufficient to label a
two-electron state by the following set of quantum numbers: the principal,
orbital, and spin quantum numbers for each electron, $n_{i}l_{i}s_{i}$
($i=1,2$), total orbital momentum $L$, total spin $S$,
total angular momentum $J$, and its projection $M$ on the quantization axis.
After the CI procedure, two-electron states may be most generally labeled
by the state energy and the quantum numbers $(L,S,J,M)$.

For singlet states $(S=0)$, the above state labeling can be simplified to
$(L,M)$, since $J$ is automatically equal to $L$.
This is not the case, however, for the triplet states $(S=1)$
since $\mathbf{J= L + S}$ due to the presence of spin-orbit interaction.
Physically, introduction of spin-orbit interactions influences the
wave functions in two aspects: The dynamical (radial) part and the
geometric (angular) part.
As for the dynamical part we neglect its influence in this paper, since
the spin-orbit interaction in the Mg atom is small \cite{aymar1}, anyway,
as one can easily see from the very small fine structure splittings,
and hence the radial wave function may be assumed to be $J-$independent
as a lowest-order approximation.
As for the geometric part, we can fully include it by introducing the
additional quantum numbers, $J$ and its projection $M$ to specify the state.
Thus it is necessary and sufficient that the triplet state is labeled
by $(L,S,J,M)$.

Now, once we have obtained the two-electron wave functions we are able
to calculate the dipole matrix elements as well as OSs for both LP and CP
fields.
In the following two subsections we present two useful conversion
relations between the $\textit{J}$-dependent and $\textit{J}$-independent
dipole matrix elements and OSs, respectively.

\subsection{Calculation of the \textit{J}-dependent dipole matrix elements }
\label{Jdmx}

By applying the well-known Wigner-Eckart theorem the following
conversion relation exists between the \textit{J}-dependent and
\textit{J}-independent dipole matrix elements if we define the
initial and final states, $i$ and $f$, by a set of quantum numbers
$\gamma_i = (n_i, L_i, S_i, J_i, M_i) $
 and
$\gamma_f = (n_f, L_f, S_f, J_f, M_f) $, respectively:

\begin{eqnarray}
D_{n_i J_i M_i n_f J_f, M_f } &=&
 		(-1)^{J_f-M_f+ L_f+S_f+ J_i+1 + q}
			 \delta_{S_i,S_f} \sqrt{(2J_i+1) (2J_f+1)}
 \nonumber \\&&
\times
\left(\begin{array}{clcr}
          J_f           &    1   &   J_i        \\
         -M_f           &    q   &   M_i    \\
        \end{array}\right)
\left\{\begin{array}{clcr}
          L_f       &    J_f   &   S_i      \\
          J_i       &    L_i   &   1        \\
        \end{array}\right\}
%\nonumber \\&&
	D_{n_i L_i M_{L_i} n_f L_f, M_{L_f}},
\label{dmx}
\end{eqnarray}

\noindent
in which $D_{n_i J_i M_i n_f J_f, M_f }$ and
$D_{n_i L_i M_{L_i} n_f L_f, M_{L_f}} $  represent the \textit{J}-dependent
and \textit{J}-independent dipole matrix elements, respectively.
$q$ is associated with laser polarization, i.e., $q=0$ for LP and
$q = \pm 1$ for right or  left circular polarization (RCP or  LCP),
respectively.
Recall that the allowed transitions take place between states accordingly
to the dipole selection rules, which are generally written as
$J_f - J_i = 0 , \pm 1$ ($J_f - J_i = 0$ is forbidden if $J_i=0$)
and $M_f - M_i = q$ ($M_f = M_i = 0$ is forbidden if $J_f - J_i = 0$).
In addition the following dipole selection rules are satisfied since
$L$ and $S$ are good quantum numbers:
$ L_f-L_i= \pm 1$, $M_{L_f}-M_{L_i}=q$  and $S_f-S_i=0$,
where $M_{L_{f(i)}}$ represents the projection on the quantization
axis of the orbital quantum momentum.

\subsection{ Calculation of the \textit{J}-dependent oscillator strengths}

Similarly, the OSs for multiplet transitions between two states, $i$ and $f$,
could be related to the \textit{J}-dependent OSs \cite{mitroy}:

\begin{equation}
f(n_i J_i,n_f J_f) = (2L_i+1) (2J_f+1)
\left\{\begin{array}{clcr}
          S_i     &    L_i   &   J_i      \\
          1       &    J_f   &   L_f      \\
        \end{array}\right\} ^2
	 f(n_i L_i, n_f L_f) ,
 \label{f}
\end{equation}

\noindent
where
$f(n_i L_i,n_f L_f)$ is the \textit{J}-independent absorption OS,
while $f(n_i J_i,n_f J_f)$ represents the \textit{J}-dependent absorption OS.

\section{Time-dependent Schr\"{o}dinger equation}
\label{se}

Having obtained the two-electron states constructed in a spherical box,
we can now solve the TDSE.
The TDSE for the two-electron atom interacting with a laser pulse reads

\begin{equation}
 i \frac{d}{dt} \Psi ({\bf r}_{1},{\bf r}_{2};t) =
 	\left[ H_{a} ({\bf r}_{1},{\bf r}_{2}) + D(t) \right]
		\Psi ({\bf r}_{1},{\bf r}_{2};t),
\label{eq:tdse}
\end{equation}

\noindent
where $ \Psi ({\bf r}_{1},{\bf r}_{2};t)$ are the total (two-electron) wave
function at positions ${\bf r}_{1} $ and $ {\bf r}_{2}$ for each electron
at time $ t $, and $ H_{a}({\bf r}_{1},{\bf r}_{2})$
is the field-free atomic Hamiltonian as shown in Eq. (\ref{eq:h_2e}).
The time-dependent interaction operator $ D(t) $ between the atom and
the laser pulse is written in the velocity gauge as,

\begin{equation}
	D(t) = - \textbf{A}(t) \cdot ({\bf p}_{1} + {\bf p}_{2}),
\label{eq:dipole}
\end{equation}

\noindent
where the dipole approximation has been employed, and
$ {\bf p}_{1} $ and $ {\bf p}_{2} $ are the momenta of the two
electrons with $ \textbf{A}(t) $ being the vector potential given by

\begin{equation}
	\textbf{A}(t)=  \textbf{A}_0 f(t)\cos (\omega t).
\label{eq:poten}
\end{equation}

\noindent
Here $ \textbf A_{0} =  {A}_{0q} \textbf{e}_q $
represents the amplitude of the vector potential and
$ \textbf{e}_q $ is the unit polarization vector of the laser pulse,
expressed in the spherical coordinates.
% with $ q = 0 $ for the LP field and $q= \pm 1$ for the RCP or  LCP field.
$\omega $ and $ f(t)$ represent the photon energy and the temporal envelope
of the laser field. In this paper we have assumed an envelope with a
cosine-squared function, i.e.,
$ f(t)=\cos ^{2}\left( {\pi t}/{2\tau }\right) $ where $ \tau $ is the
full width at half maximum (FWHM) of the vector potential  $\textbf{A}(t)$.
The integration time of Eq. (\ref{eq:tdse}) is taken from
$ -\tau $ to $ \tau $.

In order to solve Eq. (\ref{eq:tdse}), the time-dependent wave function,
$\Psi ({\bf r}_{1},{\bf r}_{2};t)$, is expanded on the atomic basis as
a linear combination of two-electron states
$\Psi ({\bf r}_{1},{\bf r}_{2};E_{n})$:

\begin{equation}
\Psi ({\bf r}_{1},{\bf r}_{2};t) =
\sum _{n  J M}C_{E_n J M}(t) \Psi ({\bf r}_{1},{\bf r}_{2};E_{n}),
\label{eq:wf_2e_t}
\end{equation}

\noindent
where $ C_{E_n J M}(t) $ is the time-dependent coefficient for a state
with an energy $ E_n $, total two-electron angular momentum $J$, and its
projection on the quantization axis $M$.
Now, by replacing Eq. (\ref{eq:wf_2e_t}) into Eq. (\ref{eq:tdse})
we obtain a set of first-order differential equations for the time-dependent
coefficients $ C_{E_n J M}(t)$:

\begin{equation}
i \frac{d}{dt} C_{E_n J M}(t) = \sum_{n',J',M'}
 	\left[ E_n \delta_{n n'} \delta_{J J'} \delta_{M M'}
	     - D_{n J M n' J' M'} (t) \right]  C_{E_{n'} J' M'}(t),
\label{eq:C-tdse}
\end{equation}

\noindent
where $ D_{n J M n' J' M'} (t) $ represents the $J$-dependent
dipole matrix element calculated in Sec. \ref{Jdmx} between two {\it triplet}
states defined by the quantum numbers $(n J M)$ and $( n' J' M')$.
This means that we have neglected the spin-forbidden transitions
between triplet and singlet states, which is reasonable for a light atom
such as Mg.
Specifically in what follows, we assume that the Mg atom is initially
in the triplet state of the lowest electronic configuration,
$3s3p$ $^3P_1$($M=0$), i.e.,

\begin{equation}
| C_{E_n J M}(t = -\tau) |^2 = \delta_{n 3} \delta_{J 1} \delta_{M 0}.
\end{equation}

\begin{figure}
\centering
\includegraphics[width=6in,angle=0]{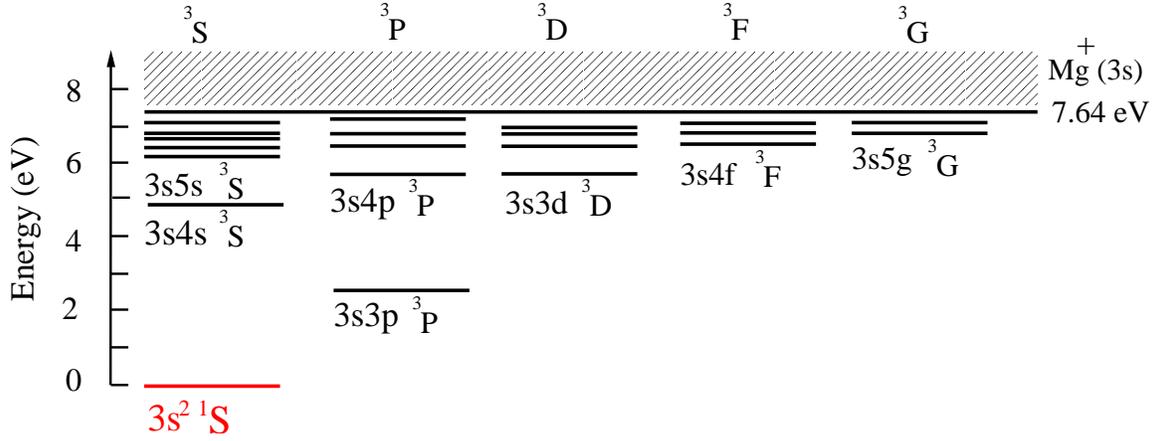}
\caption{(Color online)
Energies of the triplet states of Mg. In order to show the relative
positions with respect to the singlet ground state, $3s^2$ $^1S$ is
also shown in this energy diagram.}
\label{fig1}
\end{figure}

\begin{figure}
\centering
\includegraphics[width=3in,angle=0]{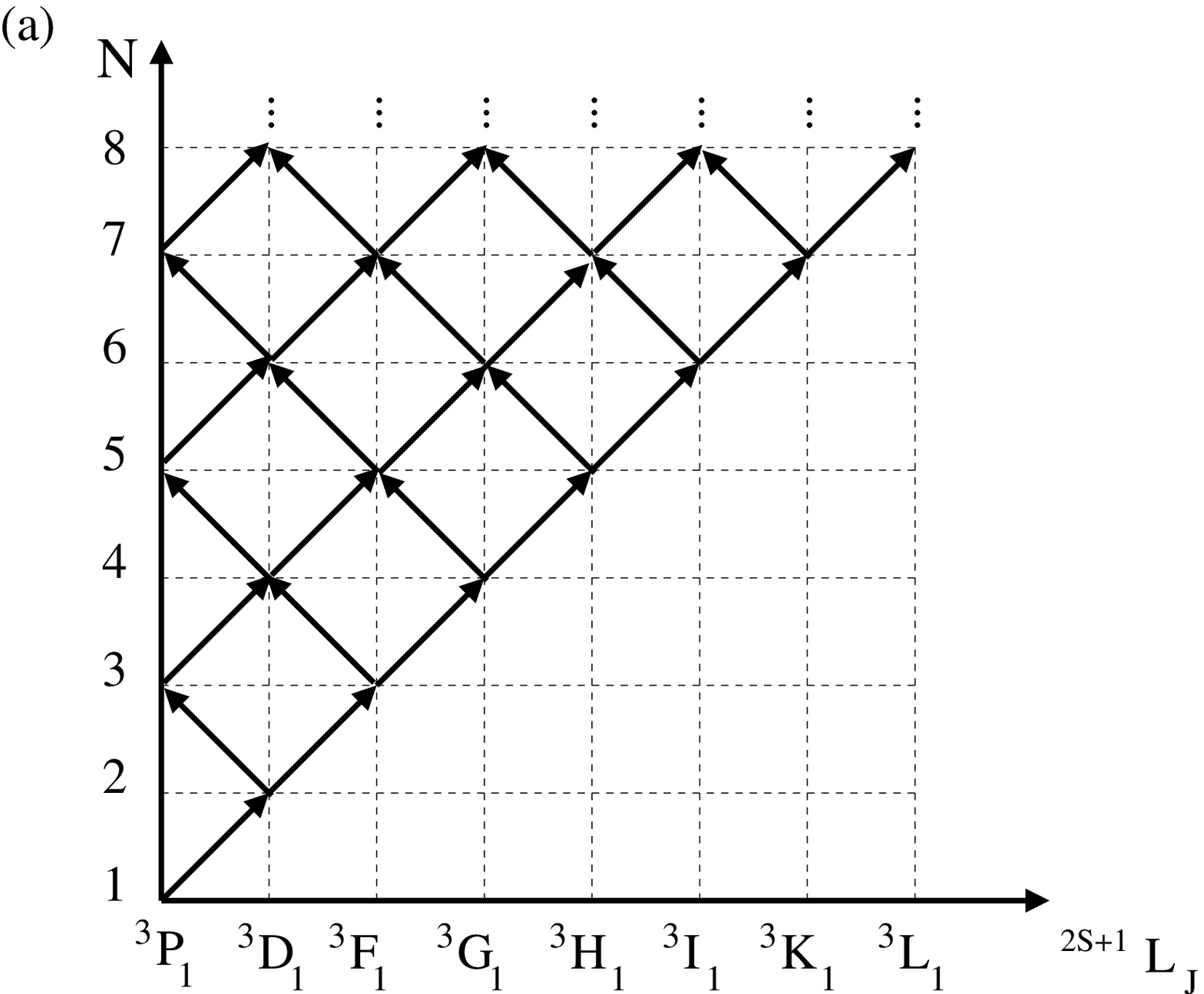}
\\
\includegraphics[width=3in,angle=0]{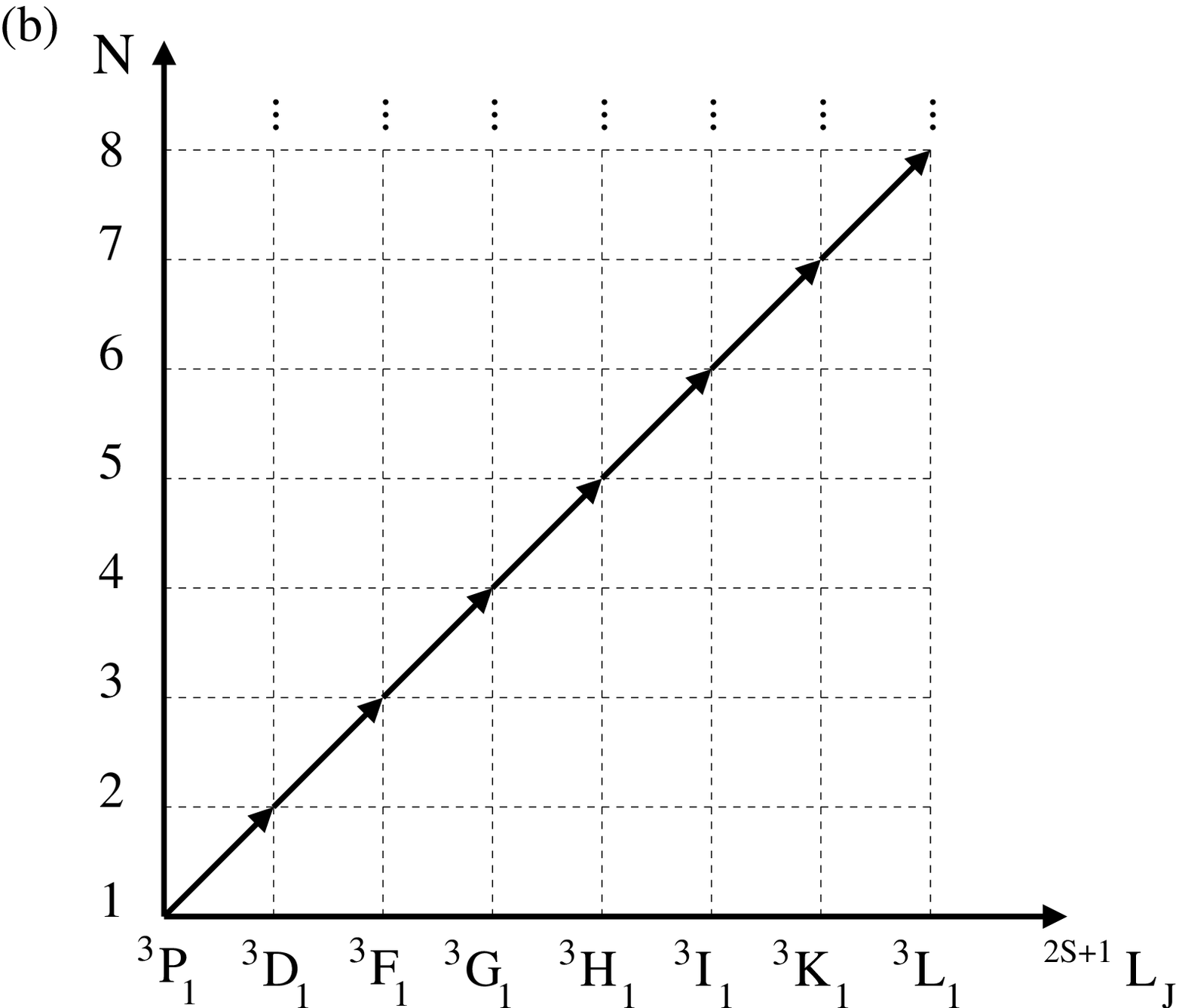}
\caption{
Allowed ionization paths in multiphoton ionization from an
$^3 P_1$ $(M=0)$ initial state by (a) a linearly polarized field
and (b)  a right circularly polarized field.}
\label{fig2}
\end{figure}

\noindent
The relevant energies of triplet states, averaged over the multiplet
components, are presented in Fig. \ref{fig1}. Note that $3s3p$ $^3P_J$
 is the lowest triplet state located at approximatively 2.71 eV from the
ground state, $3s^2$ $^1S$.
Our specific choice for the initial state results in the great simplification
of the time-dependent problem to deal with, since the allowed transition
paths by the LP or CP field become very simple as shown in
Figs. \ref{fig2}(a) and \ref{fig2}(b), respectively.
 %Recall that the dipole selection rules are generally written as
%$J - J' = 0 , \pm 1$ ($J - J' = 0$ is forbidden if $J=0$)
%and $M - M' = 0$ ($M = M' = 0$ is forbidden if $J - J' = 0$) for the LP field,
%and $J - J' = 0 , \pm 1$ and $M - M' = \pm 1$ for the RCP or  LCP field.
If we chose a different initial state, for instance, $3s4s$ $^3S_1$,
the transition paths for the LP field would be far more complicated
than those shown in Fig. \ref{fig2}(a), and accordingly
Eq. (\ref{eq:C-tdse}) would become much more difficult to solve
due to the enormous complexity of the transition paths.
In contrast, this kind of complexity does not happen for the transitions
between the singlet states \cite{gabtaka2}.

Once we have obtained the time-dependent coefficients $C_{E_n J M}$
by solving Eq. (\ref{eq:C-tdse}), the ionization yield $Y$ and
PES ${dP}/{dE}$ can be calculated at the end of the pulse:

\begin{equation}
Y= 1- \sum_{n, J, M (E_n < 0)} \mid C_{E_n J M}(t = +\tau )\mid^2,
\label{eq:yield}
\end{equation}

\noindent
 and

\begin{equation}
 \left.\frac{dP}{dE}\right\vert_{E_n = E_e} =
 \sum_{J,M (E_n = E_e)}\left|C_{E_n J M}(t=+\tau)\right|^2,
\label{eq:pes}
\end{equation}

\noindent
where $E_e$ represents the photoelectron energy.

\section{Numerical results}
\label{nr}

Before solving the TDSE we must perform several checks regarding
the accuracy of the atomic basis for the {\it triplet states} of Mg.
Related to this, we have already obtained accurate atomic basis for the
{\it singlet states} in our previous work \cite{gabtaka1} using FCHFP
as well as MP approaches.
The atomic basis states we need to solve the TDSE is constructed
in a box size of 300 a.u. for the total angular momentum up to
$J=9$ with 1000 states for each total angular momentum.
A number of $302$ \textit{B}-spline polynomials of order $9$
with a sinelike knot grid is employed.
To check the numerical convergence we have increased
the box size up to 1000 a.u. together with an increased
number of total angular momentum up to $J=14$ for each given intensity.
It turned out that the basis states constructed in a box of 300 a.u.
with the total angular momentum up to $J=9$ with 800 states for each
angular momentum are sufficient to obtain a reasonable convergence
in terms of the total ionization yield as well as PES.
In Table I we present the two-electron angular configurations of type
$(n_1 \;l_1,\; n_2 \;l_2)$ included in the construction of the two-electron
wave functions. The principal quantum numbers are taken values in the range
$  n_1 = (3- 7) $ and  $   n_2 =(1- 290)$ (with $n_1 \neq n_2$ if
$l_1 = l_2$  ), respectively for each symmetry. The number of the two-electron
 configurations varies between $1100$ and $ 1300$ for the total angular
momentum up to $J=9$.

To start with, we have compared the OSs for the {\it triplet states}
obtained by the length and velocity gauges with the FCHFP approach,
and confirmed that the agreement is quite good.
This is a good indication that our wave functions are accurate.
As for the MP approach, however, it is well known that the physically
correct dipole matrix elements can be calculated only in the length gauge
\cite{starace-kobe}, since the Hamiltonian becomes nonlocal due to
the $l$-dependence of the MP (see  Ref. \cite{gabtaka1}), and we
cannot perform a similar comparison between the two gauges.

As a more direct comparison, we have calculated the state energies, OSs,
and dipole matrix elements by both FCHFP and MP approaches and compared them
with the existing theoretical and experimental data.
In Table II we show the comparison of the calculated energies
for the first ionization threshold and the first few triplet states
$3snl$ $\;^3 L$ with the corresponding experimental values,
where $n =(3-6)$ or $(4-7)$ for each total orbital momentum $L= S,P,D$ and $F$.
The energies (in units of eV) are taken with respect to the
second ionization threshold Mg$^{2+}$ and the triplet states energies
are averaged over the multiplet components.
The theoretical data are taken from Ref. \cite{fang} and the
experimental data are taken from the database of the National
Institute of Standards and Technology (NIST) \cite{nist}.
There is an overall good agreement between the calculated energies and the
experimental values, and in addition our MP approach provides more accurate
energies than our FCHFP approach.
Of course, the accuracy of the energies do not guarantee the accuracy of
the wave functions, and we must further check the accuracy of the
wave function in terms of the \textit{J}-independent and dependent OSs.

Table III presents the comparison of the  \textit{J}-independent
OSs for single-photon transitions calculated by the
FCHFP and MP approaches with other theoretical works
Refs. \cite{fischer,moccia,chang1} and the experimental data taken from NIST.
The OSs in the length gauge are shown for single-photon
transitions among the first few triplet states:
$3s4s \;^3S \to  3s(4-7)p \;^3P $,
$3s3p \;^3P \to  3s(4-7)s \;^3S $,
$3s3p \;^3P \to  3s(3-6)d \;^3D $,
$3s3d \;^3D \to  3s(3-6)p \;^3P $,
 and
$3s3d \;^3D \to  3s(4-7)f \;^3F $.
From Table III it is clear that both FCHF and MP approaches provide
an accurate atomic basis for the triplet states of Mg, and
the overall agreement is quite well with other accurate calculations
and the experimental data.
There are, however, relatively large differences in the OSs for the
$3s4s$ $^3P \to 3s6p$ $^3P$ transition calculated by the FCHFP approach
and
$3s3d$ $^3D \to 3s4p$  $^3P$ transition calculated by the MP approach.
Besides a small difference exists in the OS of the $3s3d$ $^3D^e \to
3s4p$ $^3P^o$ transition calculated by both FCHF and MP approaches.
This might be due to the very small energy difference between these
two bound states of 0.016 eV.

Finally, in Table IV  we show the comparison of the  \textit{J}-dependent
single-photon absorption OSs calculated by the FCHF and MP approaches
with the experimental data taken from NIST.
The calculated OSs are shown for the length gauge for the single-photon
transitions among the first few triplet states:
$3s4s \;^3S_1       \;\to  3s4p \;^3P_{0,1,2} \;$,
$3s3p \;^3P_{0,1,2} \;\to  3s4s \;^3S_1       \;$,
$3s3p \;^3P_{0,1,2} \;\to  3s3d \;^3D_{1,2,3} \;$,
$3s3d \;^3D_{1,2,3} \;\to  3s5p \;^3P_{0,1,2} \;$,
and
$3s3d \;^3D_{1,2,3} \;\to  3s4f \;^3F_{2,3,4} \;$.
Again, the overall agreement is quite good between our results and the
experimental values. Therefore in what follows we present numerical TDSE
results using the the atomic basis calculated by the FCHFP only.

Having checked the accuracy of the atomic basis for the triplet states,
we are now ready to perform the time integration of Eq. (\ref{eq:C-tdse})
under various intensities for both LP and CP laser pulses.
Recall that a number of $800$ two-electron states for each total angular
momentum up to $J=9$ was used for the numerical integration of TDSE, thus
leading to a total number of $7200$ coupled differential equation to be solved.
Please note that the typical size of the dipole matrices is about $800
\times 800$.
The Runge-Kutta subroutines were used to perform the numerical integration
of TDSE.
As we have already mentioned, our initial state is $3s3p \;^3P_1$ ($M=0$)
and the photon energy is 2.7 eV which can be obtained from the second
 harmonic of a Ti:sapphire laser.
Since the energy difference from $3s3p \;^3P_1$ to the ionization threshold
is about 4.93 eV, at least two photons are needed for ionization.
The intensity range we have considered for the numerical calculations
is from $10^{11}$ W/cm$^2$ up to $10^{14}$ W/cm$^2$.
The Keldysh parameter $\gamma$ is $1.1$ at $10^{14}$ W/cm$^2$.

\begin{figure}
\centering
\includegraphics[width=4in,angle=0]{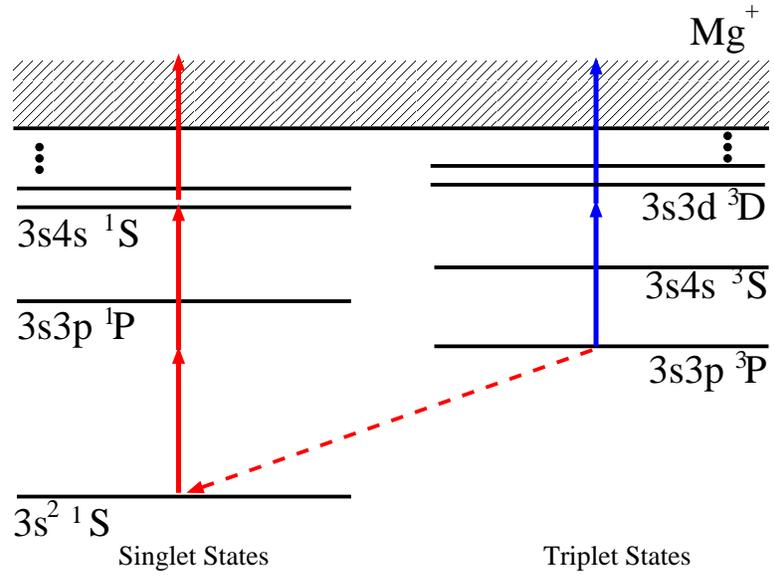}
\caption{(Color online)
Ionization scheme for Mg, at the photon energy of $ 2.7 $ eV,
which includes the atomic basis of both singlet and triplet states
resonantly coupled through the $3s^2$ $^1S$ $\to$ $3s3p$ $^3P$ transition.}
\label{fig3}
\end{figure}

The last check we should perform is that we may neglect the
entire singlet states when we solve the TDSE for the triplet states.
This check is particularly important, since our photon energy (2.7 eV)
is resonant with the spin-forbidden $3s^2 \;^1S_0 $ - $ 3s3p \;^3P_{1} $
transition.
Because we cannot calculate the dipole matrix elements for spin-forbidden
transitions within the method we use, we have taken the experimental
OS for the spin-forbidden $3s^2 \;^1S_0 $ - $ 3s3p \;^3P_{1} $ transition,
$2.38 \times 10 ^{-6}$ a.u. from NIST, which is at least five and six
 orders of magnitude smaller than those for the (nearest)
$  3s3p \;^3P \;\to 3s3d \;^3D \;$ and
$ 3s^2 \;^1S  \;\to  3s3p \;^1P \;$ transitions, respectively.
By phenomenologically including this spin-forbidden
$3s^2 \;^1S_0 $ - $ 3s3p \;^3P_{1} $ transition as shown in Fig. \ref{fig3},
we now solve two sets of TDSEs for the singlet and triplet series
which are coupled through the
resonant but very weak spin-forbidden $ 3s^2 \;^1S  \;\to  3s3p \;^1P \;$
transition.
After solving the two sets of TDSEs, we have ensured that, provided the
$ 3s3p \;^3P_{1} $ initial state, the influence of the singlet states
is extremely small as we expected, and we have safely neglected them
in the following numerical calculations.

\subsection{Ionization yield }

\begin{figure}
\centering
\includegraphics[width=4.5in,angle=0]{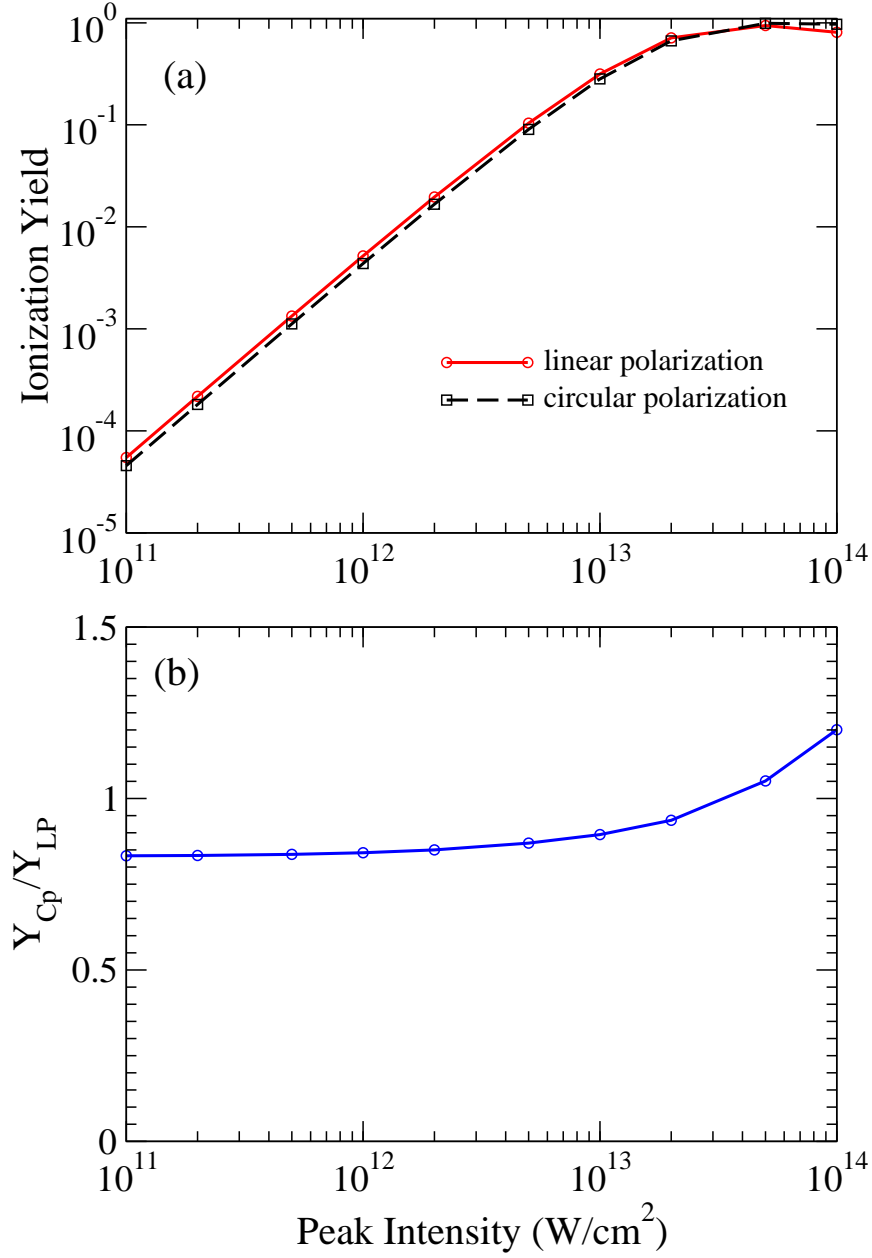}
\caption{
(Color online)
(a) Ionization yield as a function of the peak intensity for
linearly (solid) and right circularly polarized (dashed) laser
 pulses at the photon energy of  $ 2.7 $ eV.
The initial state of Mg is the triplet state $3s3p$ $^3P_1$ $(M=0)$
and the laser pulse duration is $20$ fs (FWHM).
(b) Ratio of the ionization yield by the right circularly polarized
pulse $Y_{CP}$ to that by the  linearly polarized pulse $Y_{LP}$.
}
\label{fig4}
\end{figure}

The ionization yield is shown in Fig. \ref{fig4}(a) with a log-log scale
as a function of peak intensity for the LP (solid) and RCP (dashed) pulses.
For the photon energy $2.7$ eV  we have chosen, both curves have a linear
dependence on the peak intensity, up to $ 10^{13}$ W/cm$^2$, with a slope of
1.9, indicating that our results agree well with the prediction of
lowest-order perturbation theory (LOPT).
For peak intensities higher than $2\times 10^{13}$ W/cm$^2$,
saturation starts to take place.
Figure \ref{fig4}(b) presents the ratio between the ionization
yield by the CP and LP pulses, $Y_{CP}/Y_{LP}$, as a function of
peak intensity.
For peak intensities up to $ 10^{13}$ W/cm$^2 $, the ionization
yield by the RCP pulse is about 0.83 times smaller than that by
the LP pulse.
Figures  \ref{fig4}(a) and \ref{fig4}(b) also suggest that ionization by
the LP pulse is more efficient than the RCP pulse when ionization
starts from the $3s3p$ $^3P_1$ initial state.
This result is somehow different from our previous
time-dependent calculations for multiphoton ionization of Mg
\cite{gabtaka2}: It showed that, when less than four photons are
needed for ionization, ionization from the singlet state with $^1S$ symmetry
by the CP field starts to become more efficient than that by the LP field
for a wide range of photon energy.
In the LOPT regime the main reason that the ionization yield by the CP pulse
is larger or smaller than by the LP pulse, for a non-resonant photon energy,
is determined by the particular values of the total angular momentum and its
projection on the quantization axis.

\subsection{ Photoelectron energy spectra }

In Fig. \ref{fig5} we present representative results of the PES
by the LP (solid) and RCP (dashed) pulses at the peak intensity of
$ 5 \times 10^{12}$ W/cm$^2$.
As it goes to the higher orders of above threshold ionization (ATI),
the height of the ATI peaks by the LP pulse is more than one order of
magnitude larger than that by the RCP pulse.
Of course, this could be qualitatively understood that photoionization
by the LP pulse has more chance to be near resonance with bound states
than the CP pulse, and in addition there are more accessible continua for
 the LP pulse.
It is interesting to note that subpeaks appear between
the main ATI peaks, labeled as (b) and (c), for both LP and RCP pulses,
and the height of the subpeaks is at least 5 orders of magnitude smaller
than that of the main peaks.
In addition there are small subpeaks, labeled as (a), on the right-side
shoulders of the main peaks for both LP and RCP pulses.
These results are reminiscent of the subpeaks studied in our
recent paper for the singlet states of Mg \cite{gabtaka2}, in which
multiphoton ionization of Mg from the singlet ground state has been
theoretically studied.
In that paper the origin of the subpeaks is clearly attributed to the
bound states $3snp$ $^1P$ $(n=3,4,5...)$ which are far off-resonantly
excited by the spectral wing of the pulse. In the next subsection we will
identify the origin of the subpeaks in PES in a similar manner.

Figures \ref{fig6}(a)-\ref{fig6}(c) show the variation of the PES for
three different pulse durations, (a) $\tau = 80$ fs, (b) $40$ fs, and
(c) $20$ fs (FWHM).
The photon energy and peak intensity are $ 2.7 $ eV and  $5 \times
10^{12} $ W/cm$^{2}$, respectively.
As the pulse duration decreases the ATI peaks are broadened and their
heights are decreased.
Besides, the subpeaks gradually disappear because of the broadening of the
Fourier bandwidth of the shorter pulse.

\subsection{Origin of the subpeaks in the photoelectron energy spectra}

\begin{figure}
\centering
\includegraphics[width=6in,angle=0]{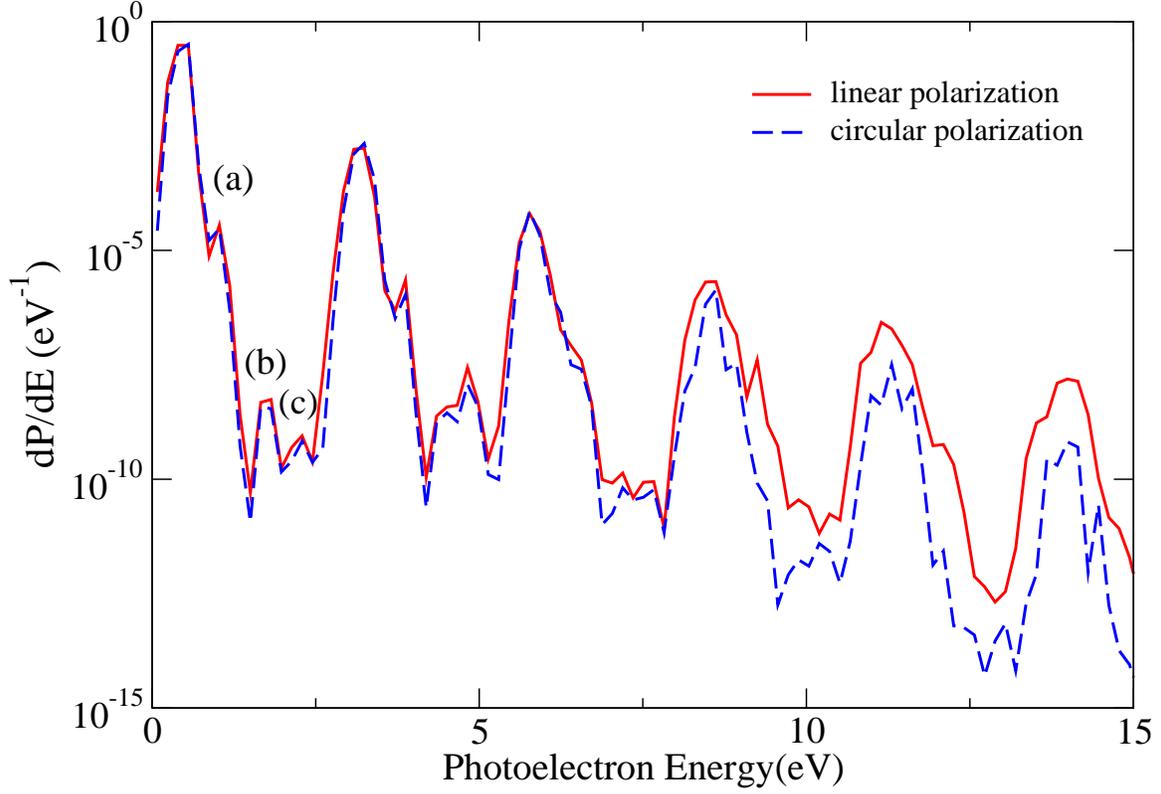}
\caption{(Color online)
 Photoelectron energy spectra by the linearly (solid) and right circularly
 polarized (dashed) laser pulses at the photon energy of $ 2.7 $ eV.
 The initial state of Mg is the triplet state $3s3p$ $^3P_1$ $(M=0)$.
The  pulse duration and peak intensity are  $20$ fs (FWHM) and $5 \times
10^{12}$ W/cm$^2$, respectively.
 }
\label{fig5}
\end{figure}

\begin{figure}
\centering
\includegraphics[width=5in,angle=0]{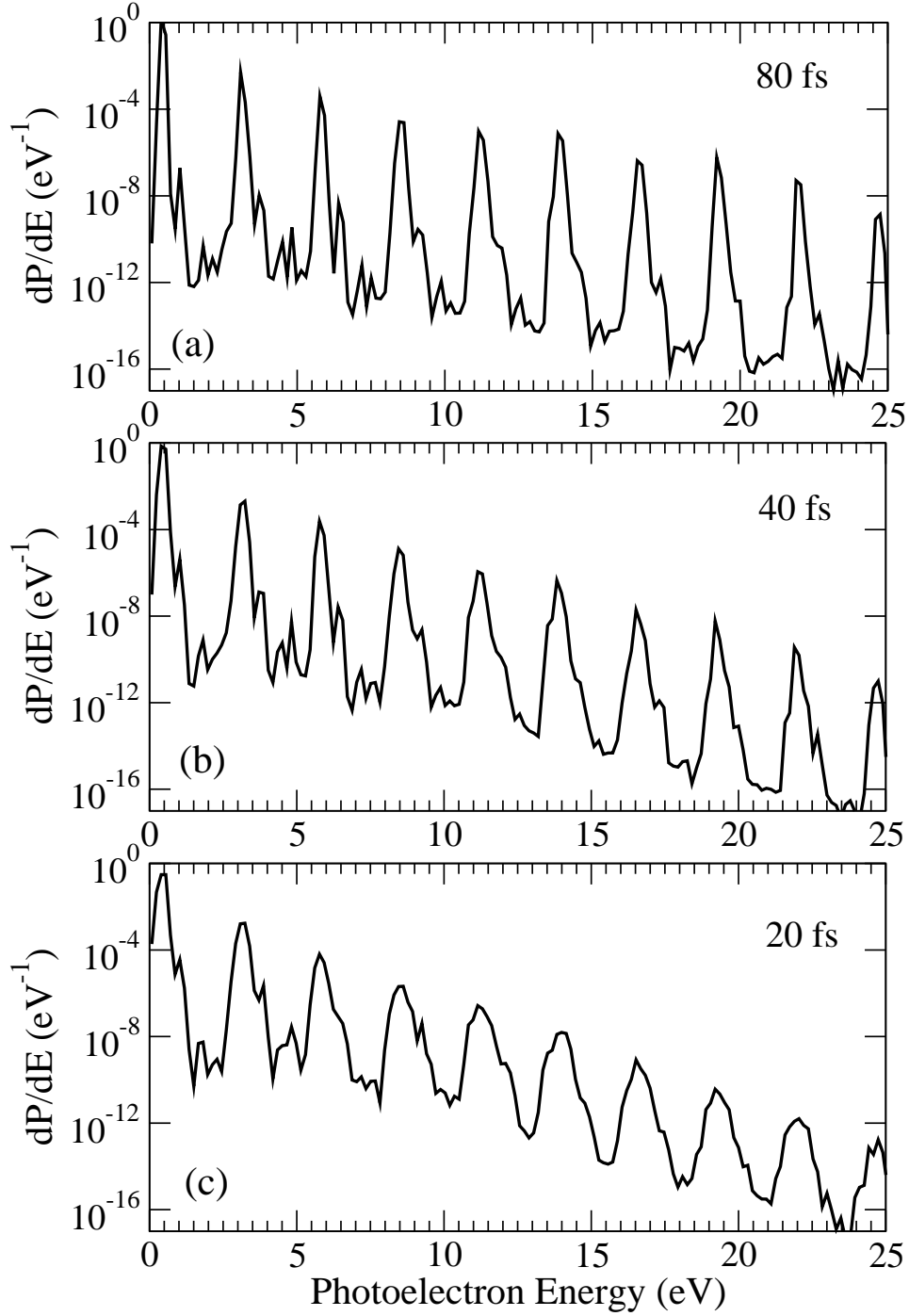}
\caption{(Color online)
Photoelectron energy spectra by the linearly laser pulses at the photon
energy of $ 2.7 $ eV and the peak intensity $5 \times 10^{12}$ W/cm$^2$ for
tree different values of the pulse duration (a) $80$ fs, (b) $40$ fs,
and (c) $20$ fs  (FWHM).
The initial state of Mg is the triplet state $3s3p$ $^3P_1$ $(M=0)$.
 }
\label{fig6}
\end{figure}

The method we have used to identify the origin of the subpeaks in the PES
mentioned in the previous subsection is quite similar to the one
employed in our previous work  \cite{gabtaka2} for singlet states of Mg:
If the subpeaks arise from some photoionization processes involving four or
five photons to leave the ionic core in some excited state, the height of the
subpeaks with respect to the main peaks would be even much smaller than those
in Fig. \ref{fig5} at the peak intensity of $5 \times 10^{12} $ W/cm$^2$,
assuming the typical excitation/ionization efficiency with four
or five photons. The subpeaks cannot be attributed to some
intensity-dependent effects, either; the ponderomotive shift is as small
as 0.098 eV at peak intensity $5 \times 10^{12} $ W/cm$^2$, and there are no
triplet states coming into resonance during the pulse duration for both LP
and CP pulses. Perhaps the subpeaks originate from the \textit{off-resonant
excitations} of some bound states, which, however, must be confirmed by the
numerical calculations.
Since we propagate the TDSE on the atomic basis, we can easily check this
by solving the TDSE after the removal of the particular bound state
under suspect, and comparing the PES with the original one with all states
included \cite{gabtaka2}.

\begin{figure}
\centering
\includegraphics[width=5in,angle=0]{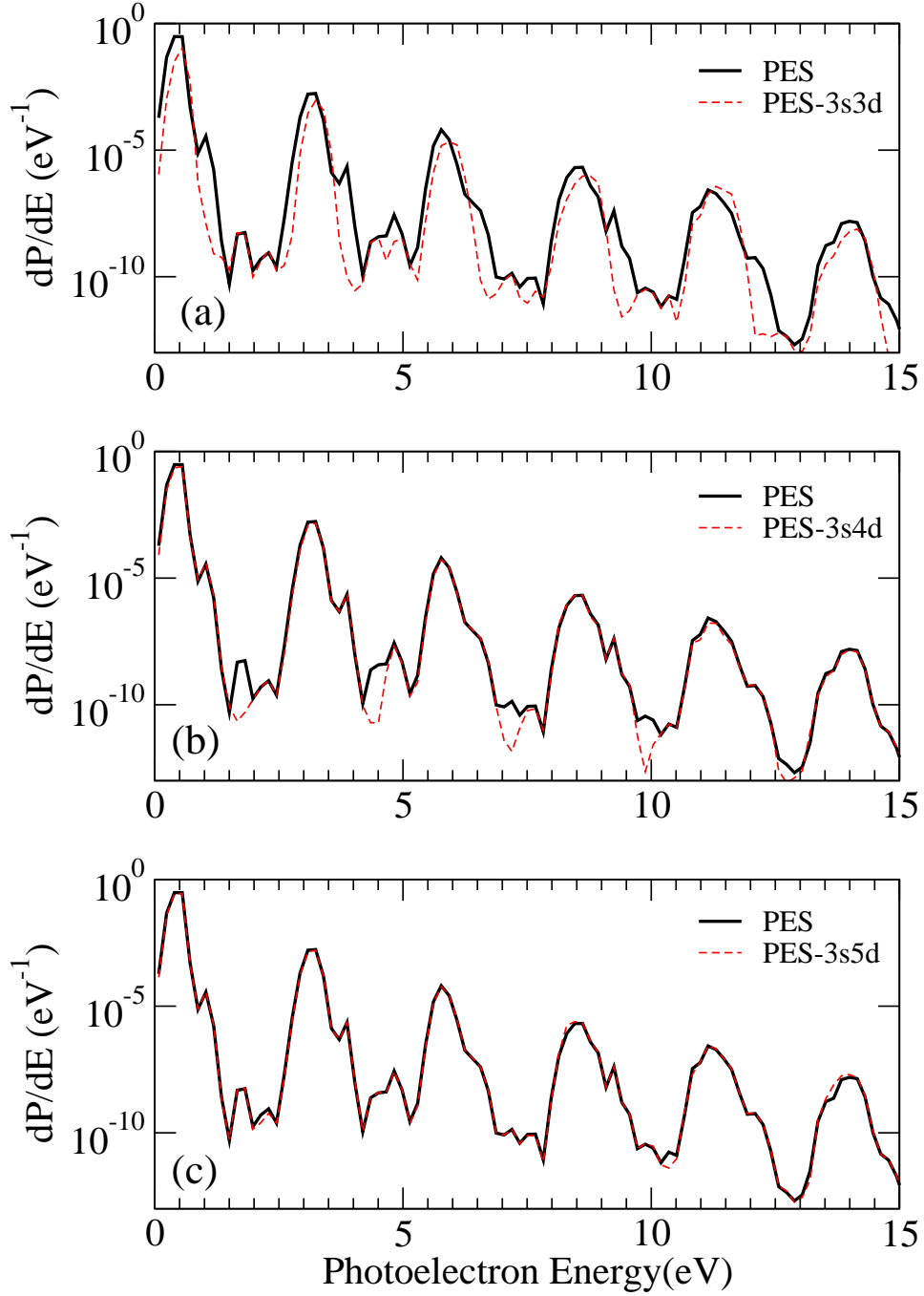}
\caption{(Color online)
Comparison of the photoelectron spectra by a linearly polarized
laser pulse  at the photon energy of $2.7 $ eV  when the (a) $3s3d$
 $^3D_1$, (b) $3s4d$ $^3D_1$, and (c) $3s5d$ $^3D_1$
 bound states of Mg are removed from the atomic basis when solving the
time-dependent Schr\"odinger equation.
The  pulse duration and peak intensity are  $20$ fs (FWHM) and $5 \times
10^{12}$ W/cm$^2$, respectively. }
\label{fig7}
\end{figure}

In Figs. \ref{fig7}(a)-\ref{fig7}(c)  we summarize the results for the PES
calculated with the LP pulse at the peak intensity of $5 \times 10^{12} $
W/cm$^2$. They are the results obtained after the removal of a
particular bound state, namely (a) $3s3d$ $^3D_1$,  (b) $3s4d$ $^3D_1$, and
(c) $3s5d$ $^3D_1$, upon solving the TDSE, and compared with the result
with the complete calculation of PES including all atomic triplet states of Mg.
When the $3s3d$ $^3D_1$ state is removed [Fig. \ref{fig7}(a)], the spike on
the right-side shoulders of each main peak disappears. In addition the height
of the main peaks is reduced since the $3s3d$ $^3D_1$ state brings an important
(but nonresonant) contribution to the ionization process.
In this particular case the laser detuning is 0.53 eV with respect
to the $3s3p$ $^3P_1$ state.
That is, the small spike, located at 1.02 eV, corresponds to the single-photon
ionization process from the off-resonantly excited $3s3d$ $^3D_1$ state.
Similarly, by removing the $3snd$ $^3D_1$ ($n=4$ and $5$) states different
subpeaks labeled as (b) and (c) in Fig. \ref{fig5} disappear, as can be seen
in Figs. \ref{fig7}(b) and \ref{fig7}(c).
This indicates that the physical origin of the subpeaks (a), (b), and (c)
in Fig. \ref{fig5} for the triplet states of Mg  is quite similar to that
we have found for the singlet states of Mg \cite{gabtaka2}.
Briefly, \textit{off-resonant} bound states such as $3snd$ $^3D_1$
($n=3,4,5$,...) are the origin of the subpeaks. Note that these states are
located at 5.94, 6.71, and 7.06 eV, respectively, from the ground state
$3s^2$ $^1S$, and accordingly the corresponding detunings are
0.53, 1.3, and 1.65 eV from the $3s3p$ $^{3}P_{1}$ state since the
photon energy is 2.7 eV.
As for Fig. \ref{fig7}(c) we note that, in addition to the $3s5d$ $^3D_1$
state, another not-identified state(s) might contribute to the subpeaks
of interest labeled as (c).

\section{Conclusions}
\label{co}

In conclusion, we have theoretically studied multiphoton ionization of Mg
from the triplet $3s3p$ $^3P_1$ state by linearly and circularly
polarized fs pulses.
For that purpose we have first constructed the atomic basis with $J-$dependent
dipole matrix elements for two active electrons, and then solved
time-dependent Schr\"{o}dinger equations with them.
Since the spin-orbit interaction is rather weak for the Mg atom,
$J-$dependent dipole matrix elements obtained by only taking into account
the geometric (angular) part of the wave functions result in rather
accurate values and compare well with the existing theoretical and
experimental data.
For the time-dependent calculations for multiphoton ionization from the
triplet $3s3p$ $^3P_1$ state, the photon energy we have specifically chosen
is 2.7 eV and corresponds to the $3s3p \;^3P_{1} \;\to  3s^2 \;^1S_0 \;$
transition which is spin-forbidden and extremely weak.
We have ensured that, even for the resonant photon energy, the singlet
states do not influence the photoionization process.
The ionization yields have been found to be larger for the linearly
polarized pulse than for the circularly polarized pulse.
Since the Mg atom has a rather rich level structure, the photoelectron energy
spectra exhibits subpeaks in addition to the ordinary main ATI peaks.
We have clarified the source of those subpeaks as ATI originating from some
triplet bound states which are far off-resonantly excited by the spectral
wing of the pulse.

\newpage
\begin{center}
\begin{tabular}{lp{5in}}
TABLE I. & Types of two-electron angular configurations used for
the construction of two-electron wave functions.
\\
\end{tabular}
\begin{tabular}{c c c c c c c c c c c c c c c c c c c c c c c c c c c c c c c }
\hline \hline
&&&  \multicolumn{1}{c}{$ {\rm ^3S^e}  $}
&&&  \multicolumn{1}{c}{$ {\rm ^3P^o}  $}
&&&  \multicolumn{1}{c}{$ {\rm ^3D^e}  $}
&&&  \multicolumn{1}{c}{$ {\rm ^3F^o}  $}
&&&  \multicolumn{1}{c}{$ {\rm ^3G^e}  $}
&&&  \multicolumn{1}{c}{$ {\rm ^3H^o}  $}
&&&  \multicolumn{1}{c}{$ {\rm ^3I^e}  $}
&&&  \multicolumn{1}{c}{$ {\rm ^3K^o}  $}
&&&  \multicolumn{1}{c}{$ {\rm ^3L^e}  $}
&&&  \multicolumn{1}{c}{$ {\rm ^3M^o}  $}
\\
\hline
&&&  \multicolumn{1}{c}{$ {\rm  ss} $}
&&&  \multicolumn{1}{c}{$ {\rm  sp}  $}
&&&  \multicolumn{1}{c}{$ {\rm  sd}  $}
&&&  \multicolumn{1}{c}{$ {\rm  sf}  $}
&&&  \multicolumn{1}{c}{$ {\rm  sg}  $}
&&&  \multicolumn{1}{c}{$ {\rm  sh}  $}
&&&  \multicolumn{1}{c}{$ {\rm  si}  $}
&&&  \multicolumn{1}{c}{$ {\rm  sk}  $}
&&&  \multicolumn{1}{c}{$ {\rm  sl}  $}
&&&  \multicolumn{1}{c}{$ {\rm  sm}  $}
\\
&&&  \multicolumn{1}{c}{$ {\rm  pp} $}
&&&  \multicolumn{1}{c}{$ {\rm  pd}  $}
&&&  \multicolumn{1}{c}{$ {\rm  pf}  $}
&&&  \multicolumn{1}{c}{$ {\rm  pd}  $}
&&&  \multicolumn{1}{c}{$ {\rm  pf}  $}
&&&  \multicolumn{1}{c}{$ {\rm  pg}  $}
&&&  \multicolumn{1}{c}{$ {\rm  ph}  $}
&&&  \multicolumn{1}{c}{$ {\rm  pi}  $}
&&&  \multicolumn{1}{c}{$ {\rm  pk}  $}
&&&  \multicolumn{1}{c}{$ {\rm  pl}  $}
\\
&&&  \multicolumn{1}{c}{$ {\rm  dd}  $}
&&&  \multicolumn{1}{c}{$ {\rm  df}  $}
&&&  \multicolumn{1}{c}{$ {\rm  dg}  $}
&&&  \multicolumn{1}{c}{$ {\rm  pg}  $}
&&&  \multicolumn{1}{c}{$ {\rm  dg}  $}
&&&  \multicolumn{1}{c}{$ {\rm  df}  $}
&&&  \multicolumn{1}{c}{$ {\rm  dg}  $}
&&&  \multicolumn{1}{c}{$ {\rm  dh}  $}
&&&  \multicolumn{1}{c}{$ {\rm  di}  $}
&&&  \multicolumn{1}{c}{$ {\rm  dk}  $}
\\
&&&  \multicolumn{1}{c}{$ {\rm  ff}  $}
&&&  \multicolumn{1}{c}{$ {\rm  fg}  $}
&&&  \multicolumn{1}{c}{$ {\rm  fh}  $}
&&&  \multicolumn{1}{c}{$ {\rm  df}  $}
&&&  \multicolumn{1}{c}{$ {\rm  fg}  $}
&&&  \multicolumn{1}{c}{$ {\rm  pi}  $}
&&&  \multicolumn{1}{c}{$ {\rm  ff}  $}
&&&  \multicolumn{1}{c}{$ {\rm  fg}  $}
&&&  \multicolumn{1}{c}{$ {\rm  gg}  $}
&&&  \multicolumn{1}{c}{$ {\rm  gh}  $}
\\
&&&  \multicolumn{1}{c}{$ {\rm  gg}  $}
&&&  \multicolumn{1}{c}{$ {\rm  gh}  $}
&&&  \multicolumn{1}{c}{$ {\rm  pp}  $}
&&&  \multicolumn{1}{c}{$ {\rm  fg}  $}
&&&  \multicolumn{1}{c}{$ {\rm  dd}  $}
&&&  \multicolumn{1}{c}{$ {\rm  fg}  $}
&&&  \multicolumn{1}{c}{$ {\rm  gg}  $}
&&&  \multicolumn{1}{c}{$ {\rm  gh}  $}
&&&  \multicolumn{1}{c}{$ {\rm  }  $}
\\
&&&  \multicolumn{1}{c}{$ {\rm  }  $}
&&&  \multicolumn{1}{c}{$ {\rm  }  $}
&&&  \multicolumn{1}{c}{$ {\rm  dd}  $}
&&&  \multicolumn{1}{c}{$ {\rm  }  $}
&&&  \multicolumn{1}{c}{$ {\rm  ff}  $}
&&&  \multicolumn{1}{c}{$ {\rm  }  $}
&&&  \multicolumn{1}{c}{$ {\rm  }  $}
&&&  \multicolumn{1}{c}{$ {\rm  }  $}
&&&  \multicolumn{1}{c}{$ {\rm  }  $}
\\
&&&  \multicolumn{1}{c}{$ {\rm  }  $}
&&&  \multicolumn{1}{c}{$ {\rm  }  $}
&&&  \multicolumn{1}{c}{$ {\rm  gg}  $}
&&&  \multicolumn{1}{c}{$ {\rm  }  $}
&&&  \multicolumn{1}{c}{$ {\rm  gg}  $}
&&&  \multicolumn{1}{c}{$ {\rm  }  $}
&&&  \multicolumn{1}{c}{$ {\rm  }  $}
&&&  \multicolumn{1}{c}{$ {\rm  }  $}
&&&  \multicolumn{1}{c}{$ {\rm  }  $}
\\
\hline\hline
\label{table}
\end{tabular}
\end{center}

\newpage
\samepage{
\begin{center}
\begin{tabular}{lp{5in}}
TABLE II. & Comparison of the energies for the first ionization threshold
and the first few triplet states of Mg.
The energies (in units of eV) are taken with respect to the second
ionization threshold Mg$^{2+}$ and the triplet state energies are averaged
over the multiplet components.
\\
\end{tabular}
\begin{tabular}{c c c c c c c c c c c c c c c c c c c c c c c c }
\hline \hline
      \multicolumn{1}{c}{$                            $}
&&&&  \multicolumn{1}{c}{$ {\rm FCHFP}                $}
&&&&  \multicolumn{1}{c}{$ {\rm MP}                   $}
&&&&  \multicolumn{1}{c}{  {\rm Theory}{\cite{fang}}   }
&&&&  \multicolumn{1}{c}{$ {\rm Exp(NIST)}            $}
\\
\hline\hline
      \multicolumn{1}{c}{$ E_{{\rm Mg}^+} $}
&&&&  \multicolumn{1}{c}{$ -15.000 $}
&&&&  \multicolumn{1}{c}{$ -15.042 $}
&&&&  \multicolumn{1}{c}{$         $}
&&&&  \multicolumn{1}{c}{$ -15.035 $}
 \\
\hline \hline
     \multicolumn{1}{c}{$ E_{3s4s} \;^3S^e  $}
&&&&  \multicolumn{1}{c}{$-17.532 $}
&&&&  \multicolumn{1}{c}{$-17.581 $}
&&&&  \multicolumn{1}{c}{$-17.578 $}
&&&&  \multicolumn{1}{c}{$-17.574 $}
 \\
\hline
  \multicolumn{1}{c}{$ E_{3s5s} \;^3S^e $}
&&&&  \multicolumn{1}{c}{$-16.212 $}
&&&&  \multicolumn{1}{c}{$-16.257 $}
&&&&  \multicolumn{1}{c}{$-16.246 $}
&&&&  \multicolumn{1}{c}{$-16.250 $}
\\
\hline
  \multicolumn{1}{c}{$ E_{3s6s}  \;^3S^e$}
&&&&  \multicolumn{1}{c}{$ -15.715$}
&&&&  \multicolumn{1}{c}{$ -15.759$}
&&&&  \multicolumn{1}{c}{$ -15.752$}
&&&&  \multicolumn{1}{c}{$ -15.752$}
 \\
\hline
  \multicolumn{1}{c}{$  E_{3s7s}  \;^3S^e$}
&&&&  \multicolumn{1}{c}{$ -15.472 $}
&&&&  \multicolumn{1}{c}{$ -15.515 $}
&&&&  \multicolumn{1}{c}{$         $}
&&&&  \multicolumn{1}{c}{$ -15.508 $}
  \\
\hline \hline
     \multicolumn{1}{c}{$ E_{3s3p} \;^3P^o$}
&&&&  \multicolumn{1}{c}{$ -19.904 $}
&&&&  \multicolumn{1}{c}{$ -19.979 $}
&&&&  \multicolumn{1}{c}{$ -20.027 $}
&&&&  \multicolumn{1}{c}{$ -19.969 $}
 \\
\hline
  \multicolumn{1}{c}{$ E_{3s4p}\;^3P^o$}
&&&&  \multicolumn{1}{c}{$ -16.708 $}
&&&&  \multicolumn{1}{c}{$ -16.756 $}
&&&&  \multicolumn{1}{c}{$ -16.756 $}
&&&&  \multicolumn{1}{c}{$ -16.749 $}
 \\
\hline
  \multicolumn{1}{c}{$  E_{3s5p} \;^3P^o $}
&&&&  \multicolumn{1}{c}{$ -15.917 $}
&&&&  \multicolumn{1}{c}{$ -15.962 $}
&&&&  \multicolumn{1}{c}{$ -15.957 $}
&&&&  \multicolumn{1}{c}{$ -15.955 $}
\\
\hline
  \multicolumn{1}{c}{$  E_{3s6p} \;^3P^o$}
&&&&  \multicolumn{1}{c}{$ -15.575 $}
&&&&  \multicolumn{1}{c}{$ -15.619 $}
&&&&  \multicolumn{1}{c}{$ -15.613 $}
&&&&  \multicolumn{1}{c}{$ -15.612 $}
  \\
\hline \hline
     \multicolumn{1}{c}{$  E_{3s3d} \;^3D^e $}
&&&&  \multicolumn{1}{c}{$-16.698 $}
&&&&  \multicolumn{1}{c}{$-16.740 $}
&&&&  \multicolumn{1}{c}{$-16.740 $}
&&&&  \multicolumn{1}{c}{$-16.736 $}
 \\
\hline
  \multicolumn{1}{c}{$  E_{3s4d} \;^3D^e$}
&&&&  \multicolumn{1}{c}{$-15.926 $}
&&&&  \multicolumn{1}{c}{$-15.969 $}
&&&&  \multicolumn{1}{c}{$-15.963 $}
&&&&  \multicolumn{1}{c}{$-15.963 $}
 \\
\hline
  \multicolumn{1}{c}{$  E_{3s5d}  \;^3D^e$}
&&&&  \multicolumn{1}{c}{$ -15.582 $}
&&&&  \multicolumn{1}{c}{$ -15.625 $}
&&&&  \multicolumn{1}{c}{$ -15.619 $}
&&&&  \multicolumn{1}{c}{$ -15.618 $}
\\
\hline
  \multicolumn{1}{c}{$  E_{3s6d} \;^3D^e$}
&&&&  \multicolumn{1}{c}{$ -15.400 $}
&&&&  \multicolumn{1}{c}{$ -15.442 $}
&&&&  \multicolumn{1}{c}{$ -15.436 $}
&&&&  \multicolumn{1}{c}{$ -15.436 $}
  \\
\hline \hline
     \multicolumn{1}{c}{$  E_{3s4f} \;^3F^o$}
&&&&  \multicolumn{1}{c}{$ -15.867$}
&&&&  \multicolumn{1}{c}{$ -15.909$}
&&&&  \multicolumn{1}{c}{$        $}
&&&&  \multicolumn{1}{c}{$ -15.903$}
 \\
\hline
  \multicolumn{1}{c}{$  E_{3s5f} \;^3F^o$}
&&&&  \multicolumn{1}{c}{$ -15.553 $}
&&&&  \multicolumn{1}{c}{$ -15.596 $}
&&&&  \multicolumn{1}{c}{$         $}
&&&&  \multicolumn{1}{c}{$ -15.589 $}
 \\
\hline
  \multicolumn{1}{c}{$  E_{3s6f} \;^3F^o$}
&&&&  \multicolumn{1}{c}{$ -15.383 $}
&&&&  \multicolumn{1}{c}{$ -15.426 $}
&&&&  \multicolumn{1}{c}{$         $}
&&&&  \multicolumn{1}{c}{$ -15.415 $}
\\
\hline
  \multicolumn{1}{c}{$  E_{3s7f} \;^3F^o$}
&&&&  \multicolumn{1}{c}{$ -15.281 $}
&&&&  \multicolumn{1}{c}{$ -15.323 $}
&&&&  \multicolumn{1}{c}{$         $}
&&&&  \multicolumn{1}{c}{$ -15.317 $}
  \\   \hline\hline\\
\label{table1}
\end{tabular}
\end{center}
}

\newpage

\begin{center}
\begin{tabular}{lp{5in}}
TABLE III. & Comparison of the \textit{J}-independent single-photon
oscillator strengths  (in a.u. and length gauge) between the first
few triplet states with $^3S^e$, $^3P^o $, $^3D^e $, and $^3F^e $ symmetry.
Numbers in square brackets indicate powers of 10.
%The experimental values of the oscillator strengths are weighted averaged of
%NIST data.
\\
\end{tabular}
\begin{tabular}{c c c c c c c c c c c c c c c}
\hline  \hline
     \multicolumn{1}{c}{$ 3s4s \;^3S^e\rightarrow$}
&&&  \multicolumn{1}{c}{$ 3s4p \;^3P^o $}
&&&  \multicolumn{1}{c}{$ 3s5p \;^3P^o $}
&&&  \multicolumn{1}{c}{$ 3s6p \;^3P^o $}
&&&  \multicolumn{1}{c}{$ 3s7p \;^3P^o $}
 \\
\hline
     \multicolumn{1}{c}{$ {\rm FCHFP} $}
&&&  \multicolumn{1}{c}{$ 1.320    $}
&&&  \multicolumn{1}{c}{$ 3.434 [-2]$}
&&&  \multicolumn{1}{c}{$ 6.963 [-3]$}
&&&  \multicolumn{1}{c}{$ 2.524 [-3]$}
\\
\hline
     \multicolumn{1}{c}{$ {\rm MP} $  }
&&&  \multicolumn{1}{c}{$  1.315     $}
&&&  \multicolumn{1}{c}{$  3.344 [-2] $}
&&&  \multicolumn{1}{c}{$  6.692 [-3] $}
&&&  \multicolumn{1}{c}{$  2.403 [-3] $}
\\
\hline
     \multicolumn{1}{c}{ {\rm Theory}\cite{moccia} }
&&&  \multicolumn{1}{c}{$  1.308    $}
&&&  \multicolumn{1}{c}{$  2.97 [-2] $}
&&&  \multicolumn{1}{c}{$  5.60 [-3] $}
&&&  \multicolumn{1}{c}{$  1.90 [-3] $}
\\
\hline
     \multicolumn{1}{c}{ {\rm Theory}\cite{fischer} }
&&&  \multicolumn{1}{c}{$  1.314    $}
&&&  \multicolumn{1}{c}{$  3.13 [-2] $}
&&&  \multicolumn{1}{c}{$  6.3  [-3] $}
&&&  \multicolumn{1}{c}{$  2.2  [-3] $}
  \\
\hline \hline
     \multicolumn{1}{c}{$ 3s3p \;^3P^o\rightarrow$}
&&&  \multicolumn{1}{c}{$ 3s4s \;^3S^e $}
&&&  \multicolumn{1}{c}{$ 3s5s \;^3S^e $}
&&&  \multicolumn{1}{c}{$ 3s6s \;^3S^e $}
&&&  \multicolumn{1}{c}{$ 3s7s \;^3S^e $}
 \\
\hline
     \multicolumn{1}{c}{$ {\rm FCHFP} $}

&&&  \multicolumn{1}{c}{$ 1.369 [-1] $}
&&&  \multicolumn{1}{c}{$ 1.546 [-2] $}
&&&  \multicolumn{1}{c}{$ 5.227 [-3] $}
&&&  \multicolumn{1}{c}{$ 2.468 [-3] $}
\\
\hline
     \multicolumn{1}{c}{$ {\rm MP }$  }
&&&  \multicolumn{1}{c}{$  1.355 [-1] $}
&&&  \multicolumn{1}{c}{$  1.533 [-2] $}
&&&  \multicolumn{1}{c}{$  5.178 [-3] $}
&&&  \multicolumn{1}{c}{$  2.442 [-3] $}
 \\
\hline
     \multicolumn{1}{c}{ {\rm Theory}\cite{fischer} }
&&&  \multicolumn{1}{c}{$  1.360 [-1]$}
&&&  \multicolumn{1}{c}{$  1.57  [-2]$}
&&&  \multicolumn{1}{c}{$  5.3   [-3]$}
&&&  \multicolumn{1}{c}{$           $}
 \\
\hline \hline
     \multicolumn{1}{c}{$ 3s3p \;^3P^o\rightarrow$}
&&&  \multicolumn{1}{c}{$ 3s3d \;^3D^e $}
&&&  \multicolumn{1}{c}{$ 3s4d \;^3D^e $}
&&&  \multicolumn{1}{c}{$ 3s5d \;^3D^e $}
&&&  \multicolumn{1}{c}{$ 3s6d \;^3D^e $}
 \\
\hline
     \multicolumn{1}{c}{$ {\rm FCHFP} $}
&&&  \multicolumn{1}{c}{$ 6.294 [-1]$}
&&&  \multicolumn{1}{c}{$ 1.263 [-1]$}
&&&  \multicolumn{1}{c}{$ 4.743 [-2]$}
&&&  \multicolumn{1}{c}{$ 2.333 [-2]$}
 \\
\hline
     \multicolumn{1}{c}{$ {\rm MP} $}
&&&  \multicolumn{1}{c}{$  6.243 [-1]  $}
&&&  \multicolumn{1}{c}{$  1.266 [-1]  $}
&&&  \multicolumn{1}{c}{$  4.772 [-2]  $}
&&&  \multicolumn{1}{c}{$  2.352 [-2]  $}

 \\
\hline
     \multicolumn{1}{c}{ {\rm Theory} \cite{fischer} }
&&&  \multicolumn{1}{c}{$  6.311 [-1]  $}
&&&  \multicolumn{1}{c}{$  1.254 [-1]  $}
&&&  \multicolumn{1}{c}{$  4.74  [-2]  $}
&&&  \multicolumn{1}{c}{$  2.32  [-2]  $}
\\
\hline \hline
     \multicolumn{1}{c}{$ 3s3d \;^3D^e\rightarrow$}
&&&  \multicolumn{1}{c}{$ 3s3p \;^3P^o $}
&&&  \multicolumn{1}{c}{$ 3s4p \;^3P^o $}
&&&  \multicolumn{1}{c}{$ 3s5p \;^3P^o $}
&&&  \multicolumn{1}{c}{$ 3s6p \;^3P^o $}
 \\
\hline
     \multicolumn{1}{c}{$ {\rm FCHFP} $}
&&&  \multicolumn{1}{c}{$ 3.776 [-1]   $}
&&&  \multicolumn{1}{c}{$ 7.520 [-3]   $}
&&&  \multicolumn{1}{c}{$ 8.551 [-3]   $}
&&&  \multicolumn{1}{c}{$ 1.714 [-3]   $}
\\
\hline
     \multicolumn{1}{c}{$ {\rm MP }   $}
&&&  \multicolumn{1}{c}{$  3.745 [-1]  $}
&&&  \multicolumn{1}{c}{$  1.014 [-2]  $}
&&&  \multicolumn{1}{c}{$  9.073 [-3]  $}
&&&  \multicolumn{1}{c}{$  1.814 [-3]  $}

\\
\hline
     \multicolumn{1}{c}{ {\rm Theory} \cite{moccia} }
&&&  \multicolumn{1}{c}{$  3.802 [-1]     $}
&&&  \multicolumn{1}{c}{$  8.0   [-3]     $}
&&&  \multicolumn{1}{c}{$  8.5   [-3]     $}
&&&  \multicolumn{1}{c}{$  1.7   [-3]     $}
\\
\hline \hline
     \multicolumn{1}{c}{$ 3s3d \;^3D^e\rightarrow$}
&&&  \multicolumn{1}{c}{$ 3s4f \;^3F^o $}
&&&  \multicolumn{1}{c}{$ 3s5f \;^3F^o $}
&&&  \multicolumn{1}{c}{$ 3s6f \;^3F^o $}
&&&  \multicolumn{1}{c}{$ 3s7f \;^3F^o $}
 \\
\hline
     \multicolumn{1}{c}{$ {\rm FCHFP} $}
&&&  \multicolumn{1}{c}{$ 7.899 [-1]$}
&&&  \multicolumn{1}{c}{$ 1.603 [-1]$}
&&&  \multicolumn{1}{c}{$ 6.017 [-2]$}
&&&  \multicolumn{1}{c}{$ 2.965 [-2]$}
\\
\hline
    \multicolumn{1}{c}{$ {\rm MP }$  }
&&&  \multicolumn{1}{c}{$  7.893 [-1] $}
&&&  \multicolumn{1}{c}{$  1.604 [-1] $}
&&&  \multicolumn{1}{c}{$  6.023 [-2] $}
&&&  \multicolumn{1}{c}{$  2.969 [-2] $}
 \\
\hline
     \multicolumn{1}{c}{ {\rm Theory} \cite{chang1} }
&&&  \multicolumn{1}{c}{$  7.97 [-1] $}
&&&  \multicolumn{1}{c}{$  1.60 [-1] $}
&&&  \multicolumn{1}{c}{$  6.00 [-2] $}
&&&  \multicolumn{1}{c}{$  2.90 [-2] $}
 \\
\hline
     \multicolumn{1}{c}{ {\rm Theory} \cite{moccia} }
&&&  \multicolumn{1}{c}{$  7.852 [-1] $}
&&&  \multicolumn{1}{c}{$  1.587 [-1] $}
&&&  \multicolumn{1}{c}{$  5.93  [-2] $}
&&&  \multicolumn{1}{c}{$  2.91  [-2] $}
 \\
\hline\hline
\label{table2}
\end{tabular}
\end{center}

\newpage

\begin{center}
\begin{tabular}{lp{5in}}
TABLE IV. & Comparison of the \textit{J}-dependent single-photon
absorption oscillator strengths (in a.u. and length gauge)
between the first few triplet states with $^3S^e_J$, $^3P^o_J $,
$^3D^{e}_{J} $, and $^3F^e_J $ symmetry.
Numbers in square brackets indicate powers of 10.
\\
\end{tabular}
\begin{tabular}{c c c c c c c c c c c c c c c}
\hline  \hline
     \multicolumn{1}{c}{$ $}
&&&  \multicolumn{1}{c}{$ {\rm FCHFP} $}
&&&  \multicolumn{1}{c}{$ {\rm MP}    $}
&&&  \multicolumn{1}{c}{$ {\rm Exp(NIST)} $}
 \\
\hline
     \multicolumn{1}{c}{$ 3s4s\;^3S^e_1 \rightarrow \;3s4p\;^3P^o_0$}
&&&  \multicolumn{1}{c}{$ 1.47 [-1]$}
&&&  \multicolumn{1}{c}{$ 1.46 [-1]$}
&&&  \multicolumn{1}{c}{$ 1.52 [-1]$}
\\
\hline
    \multicolumn{1}{c}{$ 3s4s\;^3S^e_1 \rightarrow \;3s4p\;^3P^o_1$}
&&&  \multicolumn{1}{c}{$  4.40 [-1]$}
&&&  \multicolumn{1}{c}{$  4.38 [-1]$}
&&&  \multicolumn{1}{c}{$  4.55 [-1]$}
\\
\hline
    \multicolumn{1}{c}{$ 3s4s\;^3S^e_1 \rightarrow \;3s4p\;^3P^o_2$}
&&&  \multicolumn{1}{c}{$  7.33 [-1]$}
&&&  \multicolumn{1}{c}{$  7.31 [-1]$}
&&&  \multicolumn{1}{c}{$  7.59 [-1]$}
  \\
\hline \hline
     \multicolumn{1}{c}{$ 3s3p\;^3P^o_0 \rightarrow \;3s4s\;^3S^e_1$}
&&&  \multicolumn{1}{c}{$ 1.37 [-1]$}
&&&  \multicolumn{1}{c}{$ 1.36 [-1]$}
&&&  \multicolumn{1}{c}{$ 1.35 [-1]$}
\\
\hline
    \multicolumn{1}{c}{$ 3s3p\;^3P^o_1 \rightarrow \;3s4s\;^3S^e_1$}
&&&  \multicolumn{1}{c}{$ 1.37  [-1]$}
&&&  \multicolumn{1}{c}{$ 1.36  [-1]$}
&&&  \multicolumn{1}{c}{$ 1.35  [-1]$}
\\
\hline
    \multicolumn{1}{c}{$3s3p\;^3P^o_2 \rightarrow \;3s4s\;^3S^e_1 $}
&&&  \multicolumn{1}{c}{$ 1.37  [-1]$}
&&&  \multicolumn{1}{c}{$ 1.36  [-1]$}
&&&  \multicolumn{1}{c}{$ 1.36  [-1]$}
  \\
\hline \hline
    \multicolumn{1}{c}{$ 3s3p\;^3P^o_0 \rightarrow \;3s3d\;^3D^e_1$}
&&&  \multicolumn{1}{c}{$ 6.29 [-1]$}
&&&  \multicolumn{1}{c}{$ 6.24 [-1]$}
&&&  \multicolumn{1}{c}{$ 5.93 [-1]$}
\\
\hline
    \multicolumn{1}{c}{$ 3s3p\;^3P^o_1 \rightarrow \;3s3d\;^3D^e_1$}
&&&  \multicolumn{1}{c}{$  1.57 [-1]$}
&&&  \multicolumn{1}{c}{$  1.56 [-1]$}
&&&  \multicolumn{1}{c}{$  1.48 [-1]$}
\\
\hline
    \multicolumn{1}{c}{$ 3s3p\;^3P^o_1 \rightarrow \;3s3d\;^3D^e_2$}
&&&  \multicolumn{1}{c}{$  4.72 [-1]$}
&&&  \multicolumn{1}{c}{$  4.68 [-1]$}
&&&  \multicolumn{1}{c}{$  4.45 [-1]$}
 \\
\hline
     \multicolumn{1}{c}{$ 3s3p\;^3P^o_2 \rightarrow \;3s3d\;^3D^e_1$}
&&&  \multicolumn{1}{c}{$ 6.29 [-3]$}
&&&  \multicolumn{1}{c}{$ 6.24 [-3]$}
&&&  \multicolumn{1}{c}{$ 5.94 [-3]$}
\\
\hline
     \multicolumn{1}{c}{$  3s3p\;^3P^o_2 \rightarrow \;3s3d\;^3D^e_2$}
&&&  \multicolumn{1}{c}{$  9.44 [-2]$}
&&&  \multicolumn{1}{c}{$  9.36 [-2]$}
&&&  \multicolumn{1}{c}{$  8.91 [-2]$}
\\
\hline
     \multicolumn{1}{c}{$  3s3p\;^3P^o_2 \rightarrow \;3s3d\;^3D^e_3$}
&&&  \multicolumn{1}{c}{$  5.28 [-1]$}
&&&  \multicolumn{1}{c}{$  5.24 [-1]$}
&&&  \multicolumn{1}{c}{$  4.99 [-1]$}
\\
\hline \hline
     \multicolumn{1}{c}{$ 3s3d\; ^3D^e_1 \rightarrow \;3s5p\;^3P^o_0$}
&&&  \multicolumn{1}{c}{$ 4.75 [-3]$}
&&&  \multicolumn{1}{c}{$ 5.04 [-3]$}
&&&  \multicolumn{1}{c}{$ 4.47 [-3]$}
\\
\hline
     \multicolumn{1}{c}{$  3s3d\; ^3D^e_1 \rightarrow \;3s5p\;^3P^o_1$}
&&&  \multicolumn{1}{c}{$  3.56 [-3]$}
&&&  \multicolumn{1}{c}{$  3.78 [-3]$}
&&&  \multicolumn{1}{c}{$  3.35 [-3]$}
\\
\hline
     \multicolumn{1}{c}{$  3s3d\; ^3D^e_1 \rightarrow \;3s5p\;^3P^o_2$}
&&&  \multicolumn{1}{c}{$  2.38 [-4]$}
&&&  \multicolumn{1}{c}{$  2.52 [-4]$}
&&&  \multicolumn{1}{c}{$  2.23 [-4]$}
 \\
\hline
     \multicolumn{1}{c}{$ 3s3d\; ^3D^e_2 \rightarrow \;3s5p\;^3P^o_1$}
&&&  \multicolumn{1}{c}{$ 6.41 [-3]$}
&&&  \multicolumn{1}{c}{$ 6.80 [-3]$}
&&&  \multicolumn{1}{c}{$ 6.03 [-3]$}
\\
\hline
     \multicolumn{1}{c}{$  3s3d\; ^3D^e_2 \rightarrow \;3s5p\;^3P^o_2$}
&&&  \multicolumn{1}{c}{$  2.14 [-3]$}
&&&  \multicolumn{1}{c}{$  2.26 [-3]$}
&&&  \multicolumn{1}{c}{$  2.01 [-3]$}
\\
\hline
     \multicolumn{1}{c}{$  3s3d\; ^3D^e_3 \rightarrow \;3s5p\;^3P^o_2$}
&&&  \multicolumn{1}{c}{$  8.55 [-3]$}
&&&  \multicolumn{1}{c}{$  9.07 [-3]$}
&&&  \multicolumn{1}{c}{$  8.04 [-3]$}
\\
\hline\hline
     \multicolumn{1}{c}{$  3s3d\;^3D^e_1 \rightarrow \;3s4f\;^3F^o_2$}
&&&  \multicolumn{1}{c}{$  7.90 [-1]$}
&&&  \multicolumn{1}{c}{$  7.89 [-1]$}
&&&  \multicolumn{1}{c}{$  7.76 [-1]$}
\\
\hline
     \multicolumn{1}{c}{$  3s3d\;^3D^e_2 \rightarrow \;3s4f\;^3F^o_2$}
&&&  \multicolumn{1}{c}{$  8.78 [-2]$}
&&&  \multicolumn{1}{c}{$  8.77 [-2]$}
&&&  \multicolumn{1}{c}{$  8.64 [-2]$}
\\
\hline
     \multicolumn{1}{c}{$  3s3d\;^3D^e_2 \rightarrow \;3s4f\;^3F^o_3$}
&&&  \multicolumn{1}{c}{$  7.02 [-1]$}
&&&  \multicolumn{1}{c}{$  7.02 [-1]$}
&&&  \multicolumn{1}{c}{$  6.89 [-1]$}
\\
\hline
     \multicolumn{1}{c}{$  3s3d\;^3D^e_3 \rightarrow \;3s4f\;^3F^o_2$}
&&&  \multicolumn{1}{c}{$  1.79 [-3]$}
&&&  \multicolumn{1}{c}{$  1.79 [-3]$}
&&&  \multicolumn{1}{c}{$  1.74 [-3]$}
\\
\hline
     \multicolumn{1}{c}{$  3s3d\;^3D^e_3 \rightarrow \;3s4f\;^3F^o_3$}
&&&  \multicolumn{1}{c}{$  6.26 [-2]$}
&&&  \multicolumn{1}{c}{$  6.26 [-2]$}
&&&  \multicolumn{1}{c}{$  6.17 [-2]$}
\\
\hline
     \multicolumn{1}{c}{$  3s3d\;^3D^e_3 \rightarrow \;3s4f\;^3F^o_4$}
&&&  \multicolumn{1}{c}{$  7.26 [-1]$}
&&&  \multicolumn{1}{c}{$  7.25 [-1]$}
&&&  \multicolumn{1}{c}{$  7.12 [-1]$}
\\
\hline\hline
\label{table3}
\end{tabular}
\end{center}

\end{document}